\documentclass[oldversion]{aa}
\usepackage{natbib}
\bibpunct{(}{)}{;}{a}{}{,}
\usepackage{graphicx}
\usepackage{txfonts}
\begin{document}

\title{Detection of additional Wolf-Rayet stars
in the starburst cluster Westerlund~1 with {\em SOAR}\thanks{Based on
observations made at the Southern Observatory for Astrophysical Research (Chile) and at OPD/LNA
(Brazil)}}

\author{J. H. Groh \inst{1,2} \and  A. Damineli \inst{1} \and M. Teodoro \inst{1} \and C. L.
Barbosa \inst{3}}

\institute{Instituto de Astronomia, Geof\'{i}sica e Ci\^encias Atmosf\'ericas, Universidade de
S\~ao Paulo, Rua do Mat\~ao 1226, Cidade Universit\'aria, 05508-900, S\~ao Paulo, SP, Brasil
\and Department of Physics and Astronomy, University of Pittsburgh, 3941 O'Hara Street,
Pittsburgh, PA, 15260, USA \and  IP\&D, Universidade do Vale do
Para\'{\i}ba, Av. Shihima Hifumi 2911, S\~ao Jos\'e dos Campos, 12244-000, SP,
Brasil}

\authorrunning{Groh et al.}
\titlerunning{Additional WRs in Wd~1}

\date{Received  / Accepted }

\offprints{Jose Henrique Groh, \email{groh@astro.iag.usp.br}}

\abstract
{We report the detection of 3 additional Wolf-Rayet stars in the young cluster Westerlund
1. They were selected as emission-line star candidates based on 1\,$\mu$m narrow-band imaging of
the cluster carried out at OPD/LNA (Brazil), and then  confirmed as Wolf-Rayet stars by K-band
spectroscopy performed at the 4.1\,m {\em SOAR} telescope (Chile). Together with previous works, this
increases the population of Wolf-Rayet stars detected in the cluster to 22 members. Moreover, it is
presented for the first time a K-band spectrum of the Luminous Blue Variable W243, which apparently
implies in a higher temperature than that derived from optical spectra taken in 2003.
The WC9 star WR-F was also observed, showing clear evidence of dust emission in
the K-band.

}

\maketitle

\section{\label{intro}Introduction}

Massive stars have an outstanding impact in the chemical and dynamical evolution of a galaxy, and certainly
played a key role in the reionization of the early universe \citep{loeb01}. Therefore, the precise knowledge
of how massive stars evolve in the HR diagram, and what are their physical parameters during their lifetime, is
crucial to understanding the star-formation history of the universe. Massive stars, their strong winds, their
large flux of ionizing photons, and their spectacular death as supernovae, are also the main characters
acting in harsh environments such as starburst galaxies, Wolf-Rayet (WR) galaxies and super star clusters. 

In the last decades, the starburst phenomenon and its associated massive stellar population has
been analyzed in Local Group galaxies as the LMC \citep{melnick85,walborn92,brandl96} and beyond
\citep{conti91,watson96,whitmore99,melo05}. With the tremendous development of the infrared astronomy
in the last decade, the efforts to observing massive stellar clusters turned towards our 
Galaxy \citep{figer99,figer02,figer05}, where the recognition of a massive young cluster has been
challenging due to the high reddening, uncertainties in the distance, and large spatial areas
needed to be covered by surveys.

This is probably the reason why the young stellar cluster Westerlund 1 (heenceforth Wd~1) remained in the
shadows for almost 15 years, since the pioneering spectroscopic survey of \citet{west87} suggested the
presence of a large number of early-type stars. However, after the discovery of a significant population of
WR stars in the cluster \citep{cn02}, Wd~1 has been the target of intense observations. It was revealed
that Wd~1 is very massive ($\sim~10^5~M_{\odot}$), harboring a huge population of evolved massive stars
\citep{cn05}. The last census counts for 19 WR stars \citep{nc05}, a luminous blue variable (LBV) star
\citep{cn04}, 6 yellow hypergiants \citep{cn05}, and a recently-discovered neutron star \citep{muno06}.

The goal of this work was to detect obscured emission-line stars in Wd~1. The 1\,$\mu$m
narrow-band imaging of the cluster, which was used to select the candidates, and the
follow-up K-band spectroscopy are described in Sect. \ref{obs}. In Sect.~\ref{wr} it is presented the
spectra of the newly-discovered WRs, line identifications, and spectroscopic
classification. It is shown in Sect.~\ref{other}, for the first time, the K-band spectra
of the LBV W243 and of the WC9 star WR-F. The conclusions are presented in Sect.~\ref{conc}.

\section{\label{obs}Observations}
\subsection{Imaging}
We carried out near-infrared imaging of the cluster at the 0.6\,m telescope of the OPD/LNA
(Brazil), during the nights of 2004 June 27--29. The detector used was a HAWAII HgCdTe 1024x1024
infrared array (CamIV), with a pixel size of 18.5\,$\mu$m and plate scale of 0.5''\,pixel$^{-1}$. The images were
obtained with narrow-band  filters centred approximately in the emission lines of
\ion{He}{ii}\,1.0124\,$\mu$m and \ion{He}{i}\,1.0830\,$\mu$m. The continuum was sampled with
narrow-band filters centered in 0.9900\,$\mu$m and 1.0655\,$\mu$m.  The photometric system and this
technique to detect emission-line objects was partially described by \citet{dam97}. The other
narrow-band filters used in this work (\ion{He}{i}\,1.0830\,$\mu$m and Cont2 at 1.0655\,$\mu$m)
follow a similar procedure as described by those authors. In Table~\ref{filters} we summarize the
properties of the set of filters used in this work.

\begin{table} 
\caption{Properties of the 1\,$\mu$m narrow-band filters used in this work} 
\label{filters}
\centering
\begin{tabular}{c c c}
\hline\hline
Filter & Central Wavelength ($\mu$m) & FWHM ($\mu$m) \\
\hline
Cont1 & 0.9900 & 0.0194 \\
\ion{He}{ii} & 1.0137 & 0.0088 \\
Cont2 & 1.0655 & 0.0192\\
\ion{He}{i} & 1.0825 & 0.0093 \\
\hline
\end{tabular}
\end{table}

The observing strategy was done following the standard techniques of near-infrared
imaging. For each filter, a mosaic of 4 images of 60s was obtained, dithered by 10''
in a square pattern. The frames were median-combined using a rejection
criterion to discard the stellar contribution, obtaining a mean sky image. This
sky image was subtracted from each original image, which were subsequently divided by
a normalized flat-field image. The instrumental magnitudes of the stars were
extracted using the IRAF\footnote{IRAF is distributed by the National Optical
Astronomy Observatories, which are operated by the Association of Universities for
Research in Astronomy, Inc., under cooperative agreement with the National Science
Foundation.} package  {\it DAOPHOT} \citep{stetson87}. 
The analysis was restricted to objects with errors in the instrumental
magnitude less than 0.05 mag. 

Once the instrumental magnitudes of the stars were obtained, line-continuum {\it versus} continuum diagrams
were constructed. An absolute calibration of the magnitudes was not attempted, as the aim was only to obtain a
{\it difference} in magnitude between line-continuum.
In Figure \ref{diags} it is shown this diagram for the filters centred at
\ion{He}{ii}\,1.0124\,$\mu$m and \ion{He}{i}\,1.0830\,$\mu$m. With this procedure, one can separate stars
without emission lines from stars which actually have them. 

It can clearly be seen in Fig.~\ref{diags} that 14 of the 19 WRs previously detected in Wd~1 by \citet{cn02}
and \citet{nc05} could be isolated in at least one of the filters. Those stars were labeled following the
designation of these authors, and, for clarity, hereafter it is used their identification letter. Among the 5
known WRs that were not detected using the aforementioned technique, WR-J, WR-K, and WR-R are located in
crowded regions, while WR-N lies outside the observed field. Interestingly, we could not detect any
emission from the WNVL star WR-S. This is inconsistent with the optical spectra published by \citet{nc05}
which show detectable He I emission lines. Such a behaviour can be explained if this object is an LBV
evolving towards the red side of the HR diagram. Therefore, we suggest that WR-S is an LBV candidate.

It is also shown in Fig.~\ref{diags} the {\it locii} of the LBV~W243, the YHG W12, and the sgB[e] W9. The
object W243 can barely be detected as a \ion{He}{i}\,1.0830\,$\mu$m emission-line object, which is
consistent with the equivalent width of this feature of just 5~{\AA}, measured in spectra taken in 2004 July and 2005 July
(Groh et al. 2006, in prep.). For instance, a typical WR has EW$_{\ion{He}{i}\,1.0830}\sim 300$~{\AA}. The YHG
W12 could not be resolved into its components $a$ and $b$ due to the
poor image quality of our dataset. Moreover, the star WR-J is blended with W12, which
probably explains why this YHG was detected as a \ion{He}{i}\,1.0830\,$\mu$m emission-line source in
Fig.~\ref{diags}. Alternatively, W12a itself could be an LBV going on an excursion to the blue side of the HR diagram,
thus developing \ion{He}{i} emission. To confirm this scenario, spectroscopic
and photometric observations of this object at high spatial resolution are required along the next years.

\subsection{K-band spectroscopy}

\begin{figure}
\resizebox{\hsize}{!}{\includegraphics{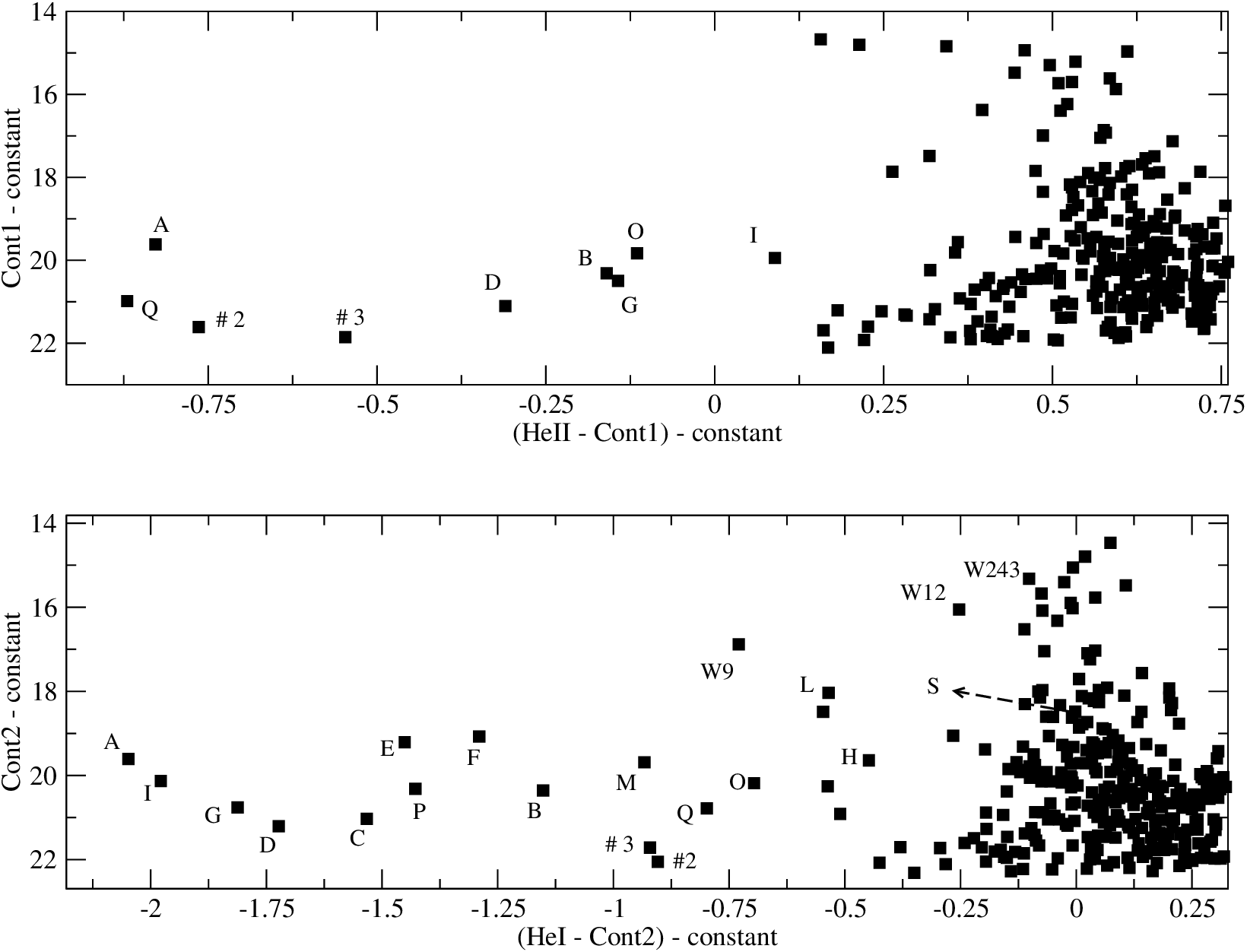}}
\caption{\label{diags}Line-continuum {\it versus} continuum diagram for the narrow-band filters used in this
work. The units of both axis are arbitrary magnitudes. {\it Upper panel:} filter centered at
\ion{He}{ii}\,1.0124\,$\mu$m. {\it Lower panel:} filter centered at \ion{He}{i}\,1.0830\,$\mu$m. 
It is noticeable, in both filters, the presence of new emission-line candidates
other than the WRs presented by \citet{nc05}; the latter are identified by their respective
identification letters.}
\end{figure}

Interestingly, new emission-line candidates in Wd~1 could be identified from the narrow-band imaging. The
spectroscopic follow-up of the candidates identified in Fig.~\ref{diags} was carried out using the {\em SOAR}
4.1\,m telescope atop Cerro Pachon, Chile. 

The near-infrared spectrograph OSIRIS\footnote{OSIRIS
(Ohio State Infrared Imager and Spectrograph) is a collaborative project between Ohio State University and CTIO.
Osiris was developed through NSF grants AST 90-16112 and AST 92-18449. OSIRIS is described in the instrument
manuals found on the CTIO Web site at http://www.ctio.noao.edu. See also \citet{depoy93}.} was used to gather long-slit
spectra of the candidates at the K-band, centered at 2.14\,$\mu$m, with R=3000. This setup provided a wavelength
coverage from 1.98 to 2.32\,$\mu$m, which allow to access the most important diagnostic lines in the K-band.
The spectra were obtained in 5 different positions of the detector, each one dithered by 5''
from the previous, in order to remove bad pixels and other detector features. The sky image was obtained by
median-combining these dithered images with a suitable rejection algorithm to remove the stellar contribution.
The mean sky frame was subtracted from each original image, which was then divided by a normalized flat-field.
The extraction of one-dimensional spectra was done using usual IRAF routines. The wavelength calibration was performed
using the sky OH lines \citep{oo92}. Telluric features were removed by dividing the extracted spectra by the
spectrum of a standard B star (HD~159402). The stellar Br\,$\gamma$ absorption feature present in HD~159402 was removed previously by
interpolating  the red and blue adjacent continuum. The individual spectra taken at different positions in the
array were then combined to a final spectrum. For each object, the final spectrum was continuum-normalized
with a low-order Legendre polynomial function.

\section{\label{wr}Newly-identified Wolf-Rayet stars}

\begin{figure}
\resizebox{\hsize}{!}{\includegraphics{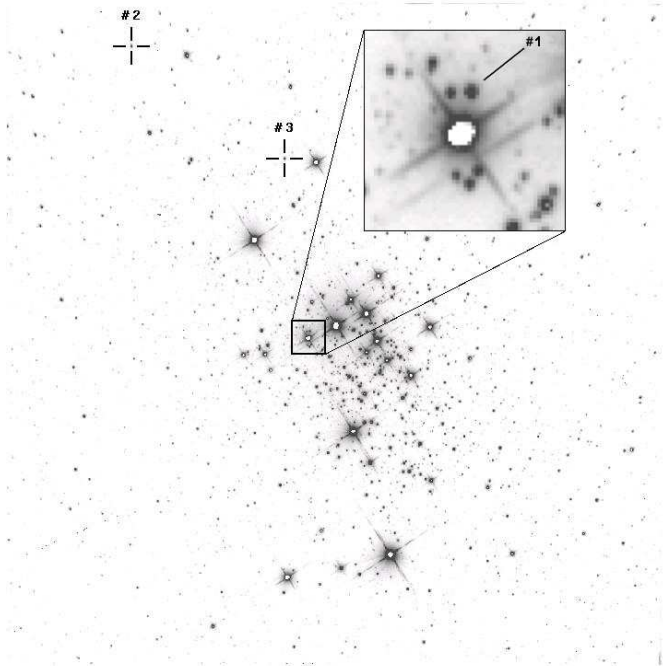}}
\caption{K-band finding chart for the new WRs discovered in Wd~1. This image was obtained with NTT/ESO.
North is up and east is to the left. The field covers a region of 4.5'x4.5'. The
inset shows a box of 15''x15'' around the star W16.\label{findchart}} 
\end{figure}

In Figure~\ref{wrs} it is displayed the K-band spectra of the newly-discovered
Wolf-Rayet stars in Wd~1 and the identification of the strongest
emission lines. The stars were classified based on the K-band spectral atlas of
WRs published by \citet{figer97}. It can be readily seen that all of them are WN stars. In 
Table~\ref{results} it is presented the basic properties of the WRs observed with the {\em SOAR} telescope, which
were named adding the prefix GDTB. A K-band finding chart is provided in Figure~\ref{findchart}.

The spectrum of GDTB~\#\,1 has a lower S/N than the other WRs shown here due to its
serendipitous detection while observing another candidate. The
presence of strong \ion{N}{v}~2.0985\,$\mu$m is ruled out, which implies in a
subtype later than WN4. Strong
lines of \ion{He}{i} are also absent, which yields a subtype earlier than WN8. Hence, GDTB~\#\,1 is 
tentatively classified as a WN5-7.

The object GDTB~\#\,2 is the earliest WN of the sample, and probably one of the earliest WNs detected in
Wd~1, rivaling with WR-A and WR-Q reported by \citet{nc05} (see also Fig. \ref{diags}). The relatively strong
emission of \ion{N}{v}\,2.0985\,$\mu$m  reveals its early type nature, together with the weakness
of the blend at 2.166\,$\mu$m (\ion{H}{i}+\ion{He}{i}+\ion{He}{ii}) compared to 
\ion{He}{ii}\, 2.1885\,$\mu$m. The precise determination of the physical parameters of GDTB~\#\,2, as of the
other WR stars present in Wd~1, relies on a quantitative spectroscopic analysis, which is beyond
the scope of this paper.

The object GDTB~\#\,3 shows a typical WN7 spectrum. While the presence of \ion{He}{i}\,2.0581\,$\mu$m is
doubtful, \ion{H}{i} is clearly present in the blending at 2.166\,$\mu$m. The other emission lines
present in the spectrum are mainly due to \ion{He}{ii} transitions. This object was detected as an x-ray source by
\citet{skinner06}.

\begin{table} 
\caption{Basic data of the newly-discovered Wolf-Rayet stars in Westerlund~1.
Coordinates and K$_S$ magnitudes were obtained from
images taken at NTT/ESO (M. Teodoro et al., in prep.).}
\label{results}
\centering
\begin{tabular}{c c c c c}
\hline\hline
Object  & $\alpha$ (J2000) & $\delta$ (J2000) & $K_{S}$ & Spectral type\\
\hline
GDTB~\#1              & 16h47m06.6s & -45$^o$50'38''.6 & 9.19 & WN5-7\\
GDTB~\#2              & 16h47m14.2s & -45$^o$48'31''.4 & 9.99 & WN4-5\\
GDTB~\#3              & 16h47m07.6s & -45$^o$49'21''.7 & 9.70 & WN7\\
\hline
\end{tabular}
\end{table} 

\begin{figure}
\resizebox{\hsize}{!}{\includegraphics{cand_02_03_84_c_identv3.eps}}
\caption{\label{wrs} Continuum-normalized spectra of the 3 new WRs detected in this work. For the sake of
clarity, the star GDTB~\#\,1 was shifted up by 2 dex and GDTB~\#\,2 by 1 dex. It can
be noticed that they are all of the WN subtype.}
\end{figure}

\section{\label{other}Other emission-line objects}

\begin{figure}
\resizebox{\hsize}{!}{\includegraphics{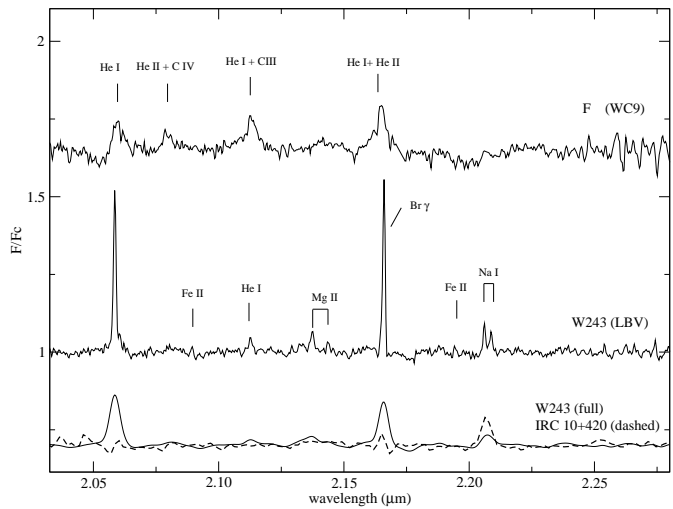}}
\caption{\label{emission} Continuum-normalized spectra of the other emission-line stars observed
in Wd~1. From top to bottom, it is shown the K-band spectrum of the WC9 star
WR-F (shifted up by 0.6 dex), of
W243 (LBV) and a comparison between the spectra of IRC~10+420 (kindly provided by Dr.~M.~Hanson) and W~243
(degraded to match the resolution of IRC 10+420).
The formers were shifted down by 0.3 dex. }
\end{figure}

As a part of the spectroscopic follow-up of the WR candidates, it was also obtained
K-band spectra of other peculiar emission-line stars in Wd~1. They are shown in Figure
\ref{emission}.

It is presented in Fig. \ref{emission} the K-band spectrum of the WC9 star WR-F, which has an optical spectrum dominated by strong \ion{C}{ii} and \ion{C}{iii}
lines \citep{cn02}. However, in its  near-infrared spectrum, WR-F shows very weak  \ion{He}{i},
\ion{He}{ii}, \ion{C}{iii} and \ion{C}{iv}
features. This is consistent with the presence of warm dust ($T\sim 1000K$) emitting on the
K-band, which veils the emission lines \citep{figer97}. Moreover, the recent claims for the presence of a
putative companion star in dusty WC9 stars \citep{williams05} would yield another source of continuum
emission, which is compatible with the low-strength nature of the emission lines of WR-F. It was also
recently detected as a moderately bright x-ray source in a {\em Chandra} observation of Wd~1 and a comparison
of its X-ray properties with other WC stars suggests that WR-F is likely a colliding-wind binary \citep{skinner05,skinner06}.

A K-band spectrum of the luminous A-supergiant W243, classified as an LBV with spectral type A2I by
\citet{cn04}, was also gathered at
the {\em SOAR} telescope. It shows strong, narrow emission lines (unresolved at our resolution of
100\,km\,s$^{-1}$) of \ion{He}{i}\,2.0581\,$\mu$m and Br\,$\gamma$. Weak emissions of 
\ion{Na}{i} $\lambda\lambda$ 2.2056--2.2084\,$\mu$m and \ion{Mg}{ii}\,$\lambda\lambda$
2.1369--2.1432\,$\mu$m doublets can also be seen. The spectral morphology at the K-band is very similar to other
LBVs, such as AG Car, LBV 1806-20 \citep{eikenberry04} and the Pistol Star \citep{figer98}. 

In Fig. \ref{emission} it is also shown a comparison between the K-band spectrum of W243 and the YHG IRC~10+420 presented
by \citet{hanson96}. It can be
noticed that IRC 10+420 has a stronger emission of \ion{Na}{i} than W243. However, 
W243 has  strong \ion{He}{i}\,2.0581\,$\mu$m and Br\,$\gamma$ emission, which are both very weak in IRC~10+420.
Hence, the K-band spectrum of W243 suggests a higher temperature than that of a typical YHG such as IRC
10+420 (T$_{eff}\leq$10kK) or than that derived from optical spectra \citep{cn04}
. Indeed, this suggestion is supported by the comparison of the K-band spectrum of W243 with supergiants
that have temperatures obtained via non-LTE models, such as AFGL 2298 (T$_{eff}=$12-15kK,
\citeauthor{clark03a}~\citeyear{clark03a}) and G26.47+0.02 (T$_{eff}=$17kK,
\citeauthor{clark03b}~\citeyear{clark03b}). Long-term spectroscopic and photometric monitoring of W243, especially in the near-infrared, is crucial
to determine its current evolutionary phase and to estimate its physical parameters.

\section{\label{conc}Concluding remarks}

In this work it was reported the discovery of 3 additional WR stars in the young starburst cluster
Westerlund~1. Their K-band spectra show that they belong to the WN subtype, in the range WN4-7. This
increases the population of WRs detected within Wd~1 from 19 \citep{nc05} to 22 members. This
work suggests that there is still a number of WRs to be discovered in this obscured super stellar
cluster, since the narrow-band imaging was performed with a modest-sized telescope (0.6\,m),
reaching a limiting magnitude of $K\simeq10.5$ mag.

It was also presented, for the first time, a K-band spectrum of a WC9 star in Wd~1
(WR-F), which clearly show evidence of continuum emission due to warm dust. 
This reduces the strength of the emission-lines in the observed spectrum, and
potentially makes it difficult to detect another putative dusty-makers WC
stars in the cluster through emission-line surveys in the K-band. Hence, a number of late WC stars may still
be waiting to be discovered in Wd~1.

The complete census of early-type stars in Wd~1 is highly desirable, as it provides an unique
``lab" to study a coeval, chemically homogeneous, large population of massive stars in the
Galaxy. Of particular interest is the precise determination of the ratio WC/WN and WR/O, since
they are dependent of the metalicity of the environment \citep{mm05}. Also, the current rate of
binarity is highly desirable to be known, especially among the massive members of the
cluster. This could give insights about the preferred mode of massive star formation, and about
the frequency of stellar mergers in such massive stellar clusters. This can likely be the case of the
peculiar sgB[e] W9, as suggested by \citet{cn05}. 

Therefore, a multi-wavelength monitoring campaign of the massive members of Wd~1 is essential over the
next years.

\begin{acknowledgements}
We thank the referee Dr. Simon Clark for valuable comments that improved the quality of the 
original manuscript. We are indebted to Steve Skinner for making available his paper before publication.
We thank Paul Crowther for insightful discussions about the identification of one of the WRs. We are grateful
to the {\em SOAR} telescope staff for obtaining the observations used in this paper during the {\em SOAR}
2005A Early Science process. We also kindly acknowledge the financial support from FAPESP (grant 02/11446-5) and CNPq (grant
200984/2004-7). J. H. Groh thanks D.~J.~Hillier and the University of Pittsburgh for partially supporting
this work.

\end{acknowledgements}

\bibliography{wr_soar_letter_v3}

\begin{thebibliography}{31}
\expandafter\ifx\csname natexlab\endcsname\relax\def\natexlab#1{#1}\fi

\bibitem[{{Brandl} {et~al.}(1996){Brandl}, {Sams}, {Bertoldi}, {Eckart},
  {Genzel}, {Drapatz}, {Hofmann}, {Loewe}, \& {Quirrenbach}}]{brandl96}
{Brandl}, B., {Sams}, B.~J., {Bertoldi}, F., {et~al.} 1996, \apj, 466, 254

\bibitem[{{Clark} {et~al.}(2003{\natexlab{a}}){Clark}, {Egan}, {Crowther},
  {Mizuno}, {Larionov}, \& {Arkharov}}]{clark03b}
{Clark}, J.~S., {Egan}, M.~P., {Crowther}, P.~A., {et~al.} 2003{\natexlab{a}},
  \aap, 412, 185

\bibitem[{{Clark} {et~al.}(2003{\natexlab{b}}){Clark}, {Larionov}, {Crowther},
  {Egan}, \& {Arkharov}}]{clark03a}
{Clark}, J.~S., {Larionov}, V.~M., {Crowther}, P.~A., {Egan}, M.~P., \&
  {Arkharov}, A. 2003{\natexlab{b}}, \aap, 403, 653

\bibitem[{{Clark} \& {Negueruela}(2002)}]{cn02}
{Clark}, J.~S. \& {Negueruela}, I. 2002, \aap, 396, L25

\bibitem[{{Clark} \& {Negueruela}(2004)}]{cn04}
{Clark}, J.~S. \& {Negueruela}, I. 2004, \aap, 413, L15

\bibitem[{{Clark} {et~al.}(2005){Clark}, {Negueruela}, {Crowther}, \&
  {Goodwin}}]{cn05}
{Clark}, J.~S., {Negueruela}, I., {Crowther}, P.~A., \& {Goodwin}, S.~P. 2005,
  \aap, 434, 949

\bibitem[{{Conti}(1991)}]{conti91}
{Conti}, P.~S. 1991, \apj, 377, 115

\bibitem[{{Damineli} {et~al.}(1997){Damineli}, {Jablonski}, {de Freitas}, \&
  {de Freitas-Pacheco}}]{dam97}
{Damineli}, A., {Jablonski}, F., {de Freitas}, L.~C., \& {de Freitas-Pacheco},
  J.~A. 1997, \pasp, 109, 633

\bibitem[{{Depoy} {et~al.}(1993){Depoy}, {Atwood}, {Byard}, {Frogel}, \&
  {O'Brien}}]{depoy93}
{Depoy}, D.~L., {Atwood}, B., {Byard}, P.~L., {Frogel}, J., \& {O'Brien}, T.~P.
  1993, in Proc. SPIE Vol. 1946, Infrared Detectors and Instrumentation, Albert
  M. Fowler; Ed., 667--672

\bibitem[{{Eikenberry} {et~al.}(2004){Eikenberry}, {Matthews}, {LaVine},
  {Garske}, {Hu}, {Jackson}, {Patel}, {Barry}, {Colonno}, {Houck}, {Wilson},
  {Corbel}, \& {Smith}}]{eikenberry04}
{Eikenberry}, S.~S., {Matthews}, K., {LaVine}, J.~L., {et~al.} 2004, \apj, 616,
  506

\bibitem[{{Figer} {et~al.}(1999){Figer}, {Kim}, {Morris}, {Serabyn}, {Rich}, \&
  {McLean}}]{figer99}
{Figer}, D.~F., {Kim}, S.~S., {Morris}, M., {et~al.} 1999, \apj, 525, 750

\bibitem[{{Figer} {et~al.}(1997){Figer}, {McLean}, \& {Najarro}}]{figer97}
{Figer}, D.~F., {McLean}, I.~S., \& {Najarro}, F. 1997, \apj, 486, 420

\bibitem[{{Figer} {et~al.}(2005){Figer}, {Najarro}, {Geballe}, {Blum}, \&
  {Kudritzki}}]{figer05}
{Figer}, D.~F., {Najarro}, F., {Geballe}, T.~R., {Blum}, R.~D., \& {Kudritzki},
  R.~P. 2005, \apjl, 622, L49

\bibitem[{{Figer} {et~al.}(2002){Figer}, {Najarro}, {Gilmore}, {Morris}, {Kim},
  {Serabyn}, {McLean}, {Gilbert}, {Graham}, {Larkin}, {Levenson}, \&
  {Teplitz}}]{figer02}
{Figer}, D.~F., {Najarro}, F., {Gilmore}, D., {et~al.} 2002, \apj, 581, 258

\bibitem[{{Figer} {et~al.}(1998){Figer}, {Najarro}, {Morris}, {McLean},
  {Geballe}, {Ghez}, \& {Langer}}]{figer98}
{Figer}, D.~F., {Najarro}, F., {Morris}, M., {et~al.} 1998, \apj, 506, 384

\bibitem[{{Hanson} {et~al.}(1996){Hanson}, {Conti}, \& {Rieke}}]{hanson96}
{Hanson}, M.~M., {Conti}, P.~S., \& {Rieke}, M.~J. 1996, \apjs, 107, 281

\bibitem[{{Loeb} \& {Barkana}(2001)}]{loeb01}
{Loeb}, A. \& {Barkana}, R. 2001, \araa, 39, 19

\bibitem[{{Melnick}(1985)}]{melnick85}
{Melnick}, J. 1985, \aap, 153, 235

\bibitem[{{Melo} {et~al.}(2005){Melo}, {Mu{\~n}oz-Tu{\~n}{\'o}n},
  {Ma{\'{\i}}z-Apell{\'a}niz}, \& {Tenorio-Tagle}}]{melo05}
{Melo}, V.~P., {Mu{\~n}oz-Tu{\~n}{\'o}n}, C., {Ma{\'{\i}}z-Apell{\'a}niz}, J.,
  \& {Tenorio-Tagle}, G. 2005, \apj, 619, 270

\bibitem[{{Meynet} \& {Maeder}(2005)}]{mm05}
{Meynet}, G. \& {Maeder}, A. 2005, \aap, 429, 581

\bibitem[{{Muno} {et~al.}(2006){Muno}, {Clark}, {Crowther}, {Dougherty}, {de
  Grijs}, {Law}, {McMillan}, {Morris}, {Negueruela}, {Pooley}, {Portegies
  Zwart}, \& {Yusef-Zadeh}}]{muno06}
{Muno}, M.~P., {Clark}, J.~S., {Crowther}, P.~A., {et~al.} 2006, \apjl, 636,
  L41

\bibitem[{{Negueruela} \& {Clark}(2005)}]{nc05}
{Negueruela}, I. \& {Clark}, J.~S. 2005, \aap, 436, 541

\bibitem[{{Oliva} \& {Origlia}(1992)}]{oo92}
{Oliva}, E. \& {Origlia}, L. 1992, \aap, 254, 466

\bibitem[{{Skinner} {et~al.}(2005){Skinner}, {Damineli}, {Palla}, {Zhekov},
  {Simmons}, \& {Teodoro}}]{skinner05}
{Skinner}, S.~L., {Damineli}, A.~D., {Palla}, F., {et~al.} 2005, BAAS, 37, 1279

\bibitem[{{Skinner} {et~al.}(2006){Skinner}, {Simmons}, {Zhekov}, {Teodoro},
  {Damineli}, \& {Palla}}]{skinner06}
{Skinner}, S.~L., {Simmons}, A.~E., {Zhekov}, S.~A., {et~al.} 2006, \apjl, 639,
  L35

\bibitem[{{Stetson}(1987)}]{stetson87}
{Stetson}, P.~B. 1987, \pasp, 99, 191

\bibitem[{{Walborn} \& {Parker}(1992)}]{walborn92}
{Walborn}, N.~R. \& {Parker}, J.~W. 1992, \apjl, 399, L87

\bibitem[{{Watson} {et~al.}(1996){Watson}, {Gallagher}, {Holtzman}, {Hester},
  {Mould}, {Ballester}, {Burrows}, {Casertano}, {Clarke}, {Crisp}, {Evans},
  {Griffiths}, {Hoessel}, {Scowen}, {Stapelfeldt}, {Trauger}, \&
  {Westphtptphal}}]{watson96}
{Watson}, A.~M., {Gallagher}, J.~S., {Holtzman}, J.~A., {et~al.} 1996, \aj,
  112, 534

\bibitem[{{Westerlund}(1987)}]{west87}
{Westerlund}, B.~E. 1987, \aaps, 70, 311

\bibitem[{{Whitmore} {et~al.}(1999){Whitmore}, {Zhang}, {Leitherer}, {Fall},
  {Schweizer}, \& {Miller}}]{whitmore99}
{Whitmore}, B.~C., {Zhang}, Q., {Leitherer}, C., {et~al.} 1999, \aj, 118, 1551

\bibitem[{{Williams} {et~al.}(2005){Williams}, {van der Hucht}, \&
  {Rauw}}]{williams05}
{Williams}, P.~M., {van der Hucht}, K.~A., \& {Rauw}, G. 2005, in Massive Stars
  and High-Energy Emission in OB Associations, ed. G.~{Rauw}, Y.~{Naz{\'e}}, \&
  R.~{Blomme}, 65--68

\end{thebibliography}

\end{document}